\documentclass[aps,prd,preprint,superscriptaddress,tightenlines,nofootinbib]{revtex4}

\usepackage{graphicx}
\usepackage{dcolumn}
\usepackage{bm}
\usepackage{epsfig}

\begin{document}

\newcommand{\ee}{e$^+$e$^-$}
\newcommand{\ff}{f$_{2}$(1525)}
\newcommand{\bb}{$b \overline{b}$}
\newcommand{\cc}{$c \overline{c}$}
\newcommand{\sbs}{$s \overline{s}$}
\newcommand{\uu}{$u \overline{u}$}
\newcommand{\dd}{$d \overline{d}$}
\newcommand{\qq}{$q \overline{q}$}
\newcommand{\suo}{\rm{\mbox{$\epsilon_{b}$}}}
\newcommand{\loro}{\rm{\mbox{$\epsilon_{c}$}}}
\newcommand{\kos}{\ifmmode \mathrm{K^{0}_{S}} \else K$^{0}_{\mathrm S} $ \fi}
\newcommand{\kol}{\ifmmode \mathrm{K^{0}_{L}} \else K$^{0}_{\mathrm L} $ \fi}
\newcommand{\ko}{\ifmmode {\mathrm K^{0}} \else K$^{0} $ \fi}

\def\tpc{three-particle correlation}
\def\twopc{two-particle correlation}
\def\ksks{K$^0_S$K$^0_S$}
\def\ee{e$^+$e$^-$}
\def\ff{f$_{2}$(1525)}

\hfill {CLNS 08/2019} 

\hfill {CLEO 08-03} 

\title{Two-Photon Widths of the $\chi_{cJ}$ States of Charmonium}

\author{K.~M.~Ecklund}
\affiliation{State University of New York at Buffalo, Buffalo, New York 14260, USA}
\author{W.~Love}
\author{V.~Savinov}
\affiliation{University of Pittsburgh, Pittsburgh, Pennsylvania 15260, USA}
\author{A.~Lopez}
\author{H.~Mendez}
\author{J.~Ramirez}
\affiliation{University of Puerto Rico, Mayaguez, Puerto Rico 00681}
\author{J.~Y.~Ge}
\author{D.~H.~Miller}
\author{I.~P.~J.~Shipsey}
\author{B.~Xin}
\affiliation{Purdue University, West Lafayette, Indiana 47907, USA}
% ...
\collaboration{CLEO Collaboration}
\noaffiliation

\date{March 19, 2008}

\begin{abstract}
Using a data sample of 24.5 million $\psi(2S)$ the reactions
$\psi(2S) \to \gamma \chi_{cJ}$, $\chi_{cJ} \to \gamma \gamma$
have been studied for the first time to determine the two-photon
widths of the $\chi_{cJ}$ states of charmonium in their decay into two
photons. The measured quantities are $\mathcal{B}(\psi(2S) \to \gamma 
\chi_{c0}) \times \mathcal{B}(\chi_{c0} \to \gamma \gamma)$=(2.22$\pm$0.32$
\pm$0.10)$\times 10^{-5}$, and $\mathcal{B}(\psi(2S) \to \gamma \chi_{c2}) 
\times \mathcal{B}(\chi_{c2}\to\gamma \gamma)$=(2.70$\pm$0.28$\pm$0.15)$\times
10^{-5}$. Using values for $\mathcal{B}(\psi(2S) \to \gamma \chi_{c0,c2})$
and $\Gamma(\chi_{c0,c2})$ from the literature the two-photon widths are 
derived to be $\Gamma_{\gamma \gamma}(\chi_{c0})$=(2.53$\pm$0.37$\pm$0.26) 
keV, $\Gamma_{\gamma \gamma}(\chi_{c2})$=(0.60$\pm$0.06$\pm$0.06) keV, and
$\mathcal{R} \equiv \Gamma_{\gamma \gamma}(\chi_{c2})$/$\Gamma_{\gamma \gamma}
(\chi_{c0})$=0.237$\pm$0.043$\pm$0.034.
The importance of the measurement of $\mathcal{R}$ is emphasized.
For the forbidden transition, $\chi_{c1} \to \gamma \gamma$, an upper limit of
$\Gamma_{\gamma \gamma}(\chi_{c1})<0.03$ keV is established.
\end{abstract}

\pacs{13.20.Gd, 13.40.Hq,14.40.Gx}
\maketitle

Charmonium spectroscopy has provided some of the most detailed information 
about the quark-antiquark interaction in Quantum Chromodynamics (QCD). 
The most practical and convenient realization of QCD for onium spectroscopy 
is in terms of perturbative QCD (pQCD), modeled
after Quantum Electrodynamics (QED). Two-photon decays of charmonium states 
$\chi_{cJ}(^{3}P_{J})$ offer the closest parallel between QED and QCD, being
completely analogous to the decays of the corresponding triplet states of
positronium. Of course, the masses of the quarks and the wave functions of the
$\chi_{c}$ states differ from those of positronium, but even these
cancel out in the ratio of the two-photon decays, so that for both positronium
and charmonium $\mathcal{R} \equiv \Gamma(^{3}P_{2} \to\gamma\gamma)/
\Gamma(^{3}P_{0}\to\gamma\gamma)$=4/15$\simeq$0.27~\cite{novikov}.
The departure from this simple lowest order prediction can arise due to 
strong radiative corrections and relativistic effects, and the measurement of 
$\mathcal{R}$ provides a unique insight into these effects. 
Two-photon decay of the spin one $\chi_{c1}$ state is forbidden by the
Landau-Yang theorem~\cite{landau}. There are numerous theoretical potential 
model predictions of $\Gamma_{\gamma\gamma}(\chi_{c0,c2})$ available in the 
literature, with some employing relativistic and/or radiative corrections.
As shown in Table~\ref{theory}, the predictions vary over a wide 
range. This underscores the importance of measuring these quantities with 
precision.

\begin{table}[h]
\begin{center}
\caption{
Potential model predictions for two-photon widths of $\chi_{c2}$ and 
$\chi_{c0}$ and the ratio $\mathcal{R}$ derived from them.
}
\begin{tabular}{lccc}
\hline \hline
Reference & $\Gamma_{\gamma\gamma}(\chi_{c2})$ (eV) & $\Gamma_{\gamma\gamma}(\chi_{c0})$ (eV) & $\mathcal{R}$ \\ \hline
Barbieri~\cite{barbieri1} & 930 & 3500 & 0.27 \\
Godfrey~\cite{godfrey} & 459 & 1290 & 0.36 \\
Barnes~\cite{barnes1} & 560 & 1560 & 0.36 \\
Bodwin~\cite{bodwin} & 820$\pm$230 & 6700$\pm$2800 & 0.12$^{+0.15}_{-0.06}$ \\
Gupta~\cite{gupta} & 570 & 6380 & 0.09 \\
M\"{u}nz~\cite{munz} & 440$\pm$140 & 1390$\pm$160 & 0.32$^{+0.16}_{-0.12}$ \\
Huang~\cite{huang} & 490$\pm$150 & 3720$\pm$1100 & 0.13$^{+0.11}_{-0.06}$ \\
Ebert~\cite{ebert} & 500 & 2900 & 0.17 \\
Schuler~\cite{schuler} & 280 & 2500 & 0.11 \\ \hline \hline
\end{tabular}
\label{theory}
\end{center}
\end{table}

Most of the existing measurements of $\Gamma_{\gamma\gamma}(\chi_{c0})$ and 
$\Gamma_{\gamma\gamma}(\chi_{c2})$ are based on 
\textit{formation} of $\chi_{cJ}$ in two-photon fusion. The only existing 
measurements based on the \textit{decay} of $\chi_{cJ}$ into two photons 
are from the Fermilab E760/E835 experiments~\cite{e8351,e8352,e8353}
with $\chi_{cJ}$ formation in $p\bar{p}$ annihilation. 
We report here results for $\Gamma_{\gamma\gamma}(\chi_{cJ})$ measured in 
the \textit{decay} of $\chi_{cJ}$ into two photons. For these measurements 
we use the reactions
\begin{equation}
\psi(2S) \to \gamma_{1}\chi_{cJ}, ~~\chi_{cJ} \to \gamma_{2}\gamma_{3},
\end{equation}
which have not been studied before. Since $\Gamma_{\gamma\gamma}(\chi_{c0})$
and $\Gamma_{\gamma\gamma}(\chi_{c2})$ are obtained from the same measurement,
we also obtain $\mathcal{R}$ with a good control of systematic errors.
Few such simultaneous measurements have been reported in the literature.

A data sample of 24.5 million $\psi(2S)$ obtained in $48~\!\mathrm{pb^{-1}}$
$e^{+}e^{-}$ annihilations at the CESR electron-positron collider was used. 
The reaction products were detected and identified using the CLEO-c detector.

The CLEO-c detector~\cite{cleodetector}, which has a cylindrical geometry, 
consists of a CsI electromagnetic calorimeter, an inner vertex drift chamber, 
a central drift chamber, and a ring-imaging Cherenkov (RICH) detector, inside 
a superconducting solenoid magnet providing a 1.0 T magnetic field. For the 
present measurements the most important component of the detector is the CsI 
calorimeter which has an acceptance of 93$\%$ of 4$\pi$ and photon energy 
resolutions of 2.2$\%$ at $E_{\gamma}$=1 GeV, and 5$\%$ at 100 MeV. 

The event selection for the final state required three photon showers, 
each with $E_{\gamma}>70$ MeV and angle $\theta$ with respect $e^{+}$ beam
direction with $|\cos\theta|<0.75$, and no charged particles. 
An energy-momentum conservation constrained 4C-fit was performed 
and events with $\chi^{2}/d.o.f.<6$ were accepted.
To prevent overlap of the lowest energy photon $\gamma_{1}$ with the high 
energy photons $\gamma_{2,3}$, events were rejected if 
$\cos\theta^{\prime}>0.98$, where $\theta^{\prime}$ is the laboratory angle 
between $\gamma_{1}$ and either $\gamma_{2}$ or $\gamma_{3}$.

Data were analyzed in two equivalent ways, by constructing the energy 
spectrum of $E(\gamma_{1})$ and the invariant mass spectrum of 
$M(\gamma_{2}\gamma_{3})$. Consistent results were obtained.
Fig.~1 shows the $E(\gamma_{1})$ spectrum.
The enhancements due to the excitation of $\chi_{c0}$ and $\chi_{c2}$ over 
substantial backgrounds are clearly observed.

In order to analyze these spectra we need to determine the shapes of the 
background and the resonance peaks. For determining peak shapes and
efficiencies fifty thousand signal Monte Carlo (MC) events were generated for 
$\chi_{c0}$ and $\chi_{c2}$ each, with
masses and widths as given by PDG 07~\cite{pdg07}. 
The radiative transition $\psi(2S) \to \gamma_{1}\chi_{c0}$ is, of course,
pure E1, and there is strong experimental evidence that the radiative 
transition $\psi(2S) \to \gamma_{1}\chi_{c2}$ is also almost pure 
E1~\cite{oreglia,mussa}.
Further, $\gamma_{2}\gamma_{3}$ in the decay $\chi_{c2} \to 
\gamma_{2}\gamma_{3}$ are expected to be produced with pure helicity two 
amplitudes~\cite{barnes1}. With these assumptions the angular distributions 
are predicted to be~\cite{kabir}
\begin{eqnarray}
&\chi_{c0}:& dN / d\cos\Theta_{1} = 1+\cos^{2}\Theta_{1}, \\
&\chi_{c2}:& d^{3}N / (d\cos\Theta_{1}d\cos\Theta_{2}d\phi_{2}) =
9\sin^{2}\Theta_{1}\sin^{2}2\Theta_{2} \nonumber \\
&&+(1+\cos^{2}\Theta_{1})[(3\cos^{2}\Theta_{2}-1)^{2}+9\sin^{4}\Theta_{2}] \nonumber \\
&&+3\sin2\Theta_{1}\sin2\Theta_{2}[3\cos^{2}\Theta_{2}-1-3\sin^{2}\Theta_{2}]\cos\phi_{2} \nonumber \\
&&+6\sin^{2}\Theta_{1}\sin^{2}\Theta_{2}(3\cos^{2}\Theta_{2}-1)\cos2\phi_{2}, \\
&\chi_{c2}:& dN/ d\cos\Theta_{1} = 1-(1/13)\cos^{2}\Theta_{1}.
\end{eqnarray}
Here $\Theta_{1}$ is the angle between $\gamma_{1}$ and the $e^{+}$ beam
direction in the $\psi(2S)$ frame, and $\Theta_{2}$ and $\phi_{2}$
are the polar and azimuthal angles of the $\gamma_{2}\gamma_{3}$ axis in 
the rest frame of $\chi_{c0,c2}$. These angular distributions were assumed 
in the MC simulations.

The energy resolutions determined by the MC simulations were 
$\sigma(E_{\gamma_{1}})$=(8.2$\pm$0.1) MeV for $\chi_{c0}$ and
$\sigma(E_{\gamma_{1}})$=(6.3$\pm$0.1) MeV for $\chi_{c2}$.
The overall efficiencies determined from 
these MC samples were $\epsilon(\chi_{c0})$=(39.1$\pm$0.5)$\%$ and 
$\epsilon(\chi_{c2})$=(50.7$\pm$0.7)$\%$. The difference between 
$\epsilon(\chi_{c0})$ and $\epsilon(\chi_{c2})$ arises primarily from the
$\cos\Theta_{1}$ distributions (Eqs.~(2) and (4)). 

\begin{figure}[h]
\begin{center}
\includegraphics[width=3.1in]{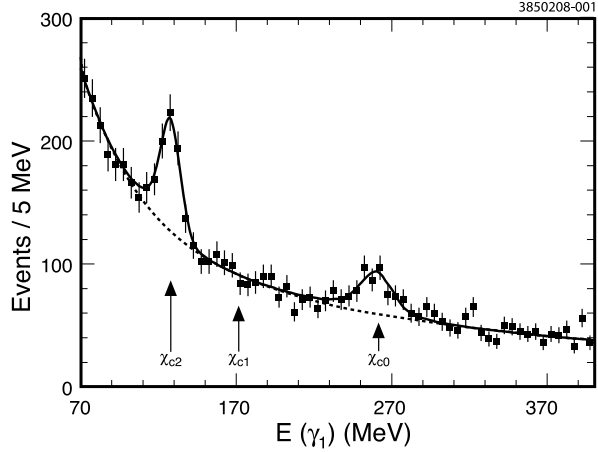}
\end{center}
\caption{Fitted spectrum for $E(\gamma_{1})$. The expected positions of
$E(\gamma_{1})$ from $\chi_{c0}$, $\chi_{c1}$, $\chi_{c2}$ are marked with
arrows.
}
\end{figure}

\begin{figure}[h]
\begin{center}
\includegraphics[width=3.1in,height=2.3in]{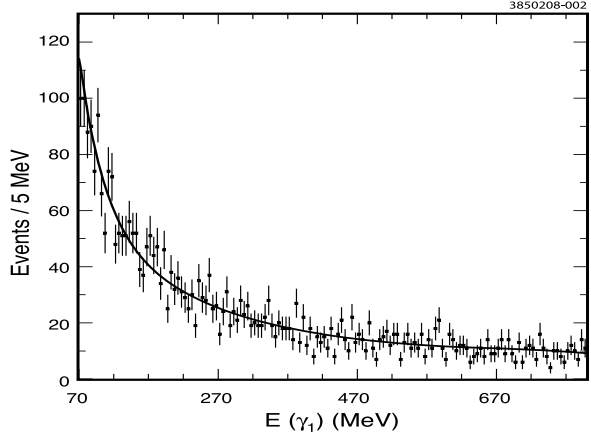}
\end{center}
\caption{Background spectrum for $E(\gamma_{1})$.
The points are from the off-$\psi(2S)$ data.
The curve is from fit to the $\psi(3770)$ data.
}
\end{figure}

Because the background in our spectrum is large, it is important to determine 
its shape accurately.
For this purpose the distributions of $E(\gamma_{1})$ were examined in the
$21~\!\mathrm{pb^{-1}}$ of off-$\psi(2S)$ data taken at $\sqrt{s}$=3671 MeV, 
as well as the $280~\!\mathrm{pb^{-1}}$ of large statistics $\psi(3770)$ data 
taken at $\sqrt{s}$=3772 MeV. As shown in Fig.~2, it is found that the 
off-$\psi(2S)$ data are in excellent agreement
with the high statistics $\psi(3770)$ data, in which transitions to either
$\chi_{c0}$ or $\chi_{c2}$ resonances were expected to yield 
$\leq$2 events~\cite{psi3770chi}.
The $E(\gamma_{1})$ distribution for the $\psi(3770)$ data was  
fitted with a polynomial, and used as the shape of the background in the 
$\psi(2S)$ data shown in Fig.~1.

The final fit to the $E(\gamma_{1})$ spectrum obtained with fixed 
$\chi_{c0}$ and $\chi_{c2}$ masses and intrinsic widths~\cite{pdg07}  
is shown in Fig.~1. The $\chi^{2}/d.o.f.=41/61$ for the fit.
It was found that $N(\chi_{c0})$=212$\pm$31 and $N(\chi_{c2})$=335$\pm$35.
The product branching ratios were determined as
$N(\chi_{cJ})/[\epsilon(\chi_{cJ})\times N(\psi(2S)]$ with the results
\begin{equation}
\begin{array}{c}
\mathcal{B}(\psi(2S) \to \gamma \chi_{c0}) \times \mathcal{B}(\chi_{c0} \to \gamma \gamma) \\
=(2.22 \pm 0.32(\mathrm{stat}))\times 10^{-5}, \\
\mathcal{B}(\psi(2S) \to \gamma \chi_{c2}) \times \mathcal{B}(\chi_{c2} \to \gamma \gamma) \\
=(2.70 \pm 0.28(\mathrm{stat}))\times 10^{-5}. \\
\end{array}
\end{equation}

Various possible sources of systematic errors in our results were 
investigated.
The number of $\psi(2S)$ produced is determined using the
background-subtracted and efficiency-corrected yield of hadronic
events following the procedure described in detail in~\cite{crball}.
The background is estimated using the off-$\psi(2S)$
data.  The efficiency is estimated by a MC simulation of generic 
$\psi(2S)$ decays. The systematic uncertainty is determined by varying
the hadronic event selection and online trigger criteria by large amounts 
in both data and MC. It is found that while the MC determined efficiency 
changes from 65\% to 91\% the efficiency-corrected yield changes by no more 
than 2\%, which we include as a systematic error.
The neutral trigger efficiency is uncertain by 0.2$\%$. 
The uncertainty in our MC determination of absolute efficiency for 
three-photon detection was estimated as $3\times 0.4\%$=1.2$\%$~\cite{photsys}.
The systematic error due to the simulation of the event selection criteria 
($\chi^{2}/d.o.f.$ distribution for 4C fit, acceptance variation, and shower
overlap cut) was
determined by varying the cuts. Similarly, systematic uncertainties due to our 
choice of the background and signal shapes were estimated by using a free 
parameter polynomial background shape and a free parameter Crystal Ball line 
shape~\cite{crball} convoluted with appropriate Breit-Wigner resonance shapes 
for the peaks. The extreme changes in the resonance yields obtained with 
these changes were taken as measures of systematic errors.
We have assumed pure helicity two decay of $\chi_{c2}$. In a relativistic
calculation Barnes~\cite{barnes1} predicts the helicity zero component
to be only 0.5$\%$. To be very conservative, we have determined the change 
in our result for $\chi_{c2}$ by including a helicity zero component of 8$\%$, 
which is the experimental upper limit established by CELLO~\cite{angular} 
for the two photon decay of the $2^{++}$ light quark state $a_{2}(1320)$.
All individual systematic errors are listed in Table~\ref{systematics}. 
The sums of the systematic errors, added in quadrature are $\pm$4.5$\%$ 
for $\chi_{c0}$ and $\pm$5.7$\%$ for $\chi_{c2}$.

\begin{table}[h]
\begin{center}
\caption{
Estimates of systematic uncertainties. Asterisks denote the systematic
sources common to both $\chi_{c0}$ and $\chi_{c2}$.
}
\begin{tabular}{lcc}
\hline \hline
Source of Systematic Uncertainty & $\chi_{c0}$ & $\chi_{c2}$  \\ \hline
Number of $\psi(2S)$$^{*}$ & 2.0$\%$ & 2.0$\%$ \\
Neutral Trigger Efficiency$^{*}$ & 0.2$\%$ & 0.2$\%$ \\
Photon Detection Efficiency $^{*}$ & 1.2$\%$ & 1.2$\%$ \\
Event Selection Simulation & 2.0$\%$ & 2.0$\%$ \\
Resonance Fitting & 3.3$\%$ & 4.6$\%$ \\
Helicity 2 Angular Distribution & - & 1.3$\%$ \\ \hline
Sum in quadrature & 4.5$\%$ & 5.7$\%$ \\ \hline \hline
\end{tabular}
\label{systematics}
\end{center}
\end{table}

Our final results for the measured quantities, $\mathcal{B}(\psi(2S) \to 
\gamma \chi_{c0,c2}) \times \mathcal{B}(\chi_{c0,c2} \to \gamma \gamma)$ 
are presented in Table~\ref{results}. We use the PDG 07 average results, 
\begin{eqnarray}
\mathcal{B}(\psi(2S) \to \gamma \chi_{c0})&=&(9.2\pm0.4)\times 10^{-2}, \nonumber \\
\Gamma(\chi_{c0})&=&(10.5\pm0.9)~\mathrm{MeV}, \nonumber \\
\mathcal{B}(\psi(2S) \to \gamma \chi_{c2})&=&(8.8\pm0.5)\times 10^{-2}, \nonumber \\
\Gamma(\chi_{c2})&=&(1.95\pm0.13)~\mathrm{MeV}, 
\end{eqnarray}
to derive $\mathcal{B}(\chi_{c0,c2} \to \gamma \gamma)$,  
$\Gamma_{\gamma\gamma}(\chi_{c0,c2})$, and 
$\mathcal{R}$. These are also listed in Table~\ref{results}.

\begin{table}[h]
\begin{center}
\caption{Results of the present measurements. First error is statistical,
second is systematic, and third is due to the PDG parameters used.
The common systematic errors have been removed in calculating $\mathcal{R}$.
$\mathcal{B}_{1}\equiv\mathcal{B}(\psi(2S) \to \gamma \chi_{c0,c2})$,
$\mathcal{B}_{2}\equiv\mathcal{B}(\chi_{c0,c2} \to \gamma\gamma)$,
$\Gamma_{\gamma\gamma}\equiv\Gamma_{\gamma\gamma}(\chi_{c0,c2} \to 
\gamma\gamma)$.
}
\begin{tabular}{lcc}
\hline \hline
Quantity & $\chi_{c0}$ & $\chi_{c2}$ \\ \hline
$\mathcal{B}_{1}\times\mathcal{B}_{2}\times 10^{5}$ & 2.22$\pm$0.32$\pm$0.10 & 2.70$\pm$0.28$\pm$0.15 \\
$\mathcal{B}_{2} \times 10^{4}$ & 2.41$\pm$0.35$\pm$0.11$\pm$0.10 & 3.06$\pm$0.32$\pm$0.17$\pm$0.17 \\
$\Gamma_{\gamma\gamma}$ (keV) & 2.53$\pm$0.37$\pm$0.11$\pm$0.24 & 0.60$\pm$0.06$\pm$0.03$\pm$0.05 \\
$\mathcal{R}$ & \multicolumn{2}{c}{0.237$\pm$0.043(stat)$\pm$0.015(syst)$\pm$0.031(PDG)} \\
\hline \hline
\end{tabular}
\label{results}
\end{center}
\end{table}

\begin{table*}[t]
\begin{center}
\caption{Compilation of experimental results for two-photon partial widths of 
$\chi_{c0}$ and $\chi_{c2}$.
}
\begin{tabular}{lcccc}
\hline \hline
Experiment [Ref.] & Measured & $\Gamma_{\gamma\gamma}(\chi_{c0})$ keV$^{*}$ & $\Gamma_{\gamma\gamma}(\chi_{c2})$ keV$^{*}$ & $\mathcal{R}$ \\ \hline
E760(1993)~\cite{e8351} & $\mathcal{B}(\bar{p}p \to \chi_{c2})\times \mathcal{B}_{\gamma\gamma}$ & - & 0.47$\pm$0.12$\pm$0.07 & - \\
E835(2000)~\cite{e8352} & $\mathcal{B}(\bar{p}p \to \chi_{cJ})\times \mathcal{B}_{\gamma \gamma}$ & 2.01$\pm$1.03$\pm$0.24 & 0.39$\pm$0.07$\pm$0.03 & 0.20$\pm$0.11$\pm$0.03 \\
E835(2004)~\cite{e8353} & $\mathcal{B}(\bar{p}p \to \chi_{c0})\times \mathcal{B}_{\gamma \gamma}$ & 3.3$\pm$0.6$\pm$0.5 & - & - \\
& & & & \\
OPAL(1998)~\cite{opal} & $\Gamma_{\gamma\gamma}\times \mathcal{B}(\chi_{c2} \to \gamma J/\psi)$ & - & 1.19$\pm$0.32$\pm$0.26 & - \\
L3(1999)~\cite{l3} & $\Gamma_{\gamma\gamma}\times \mathcal{B}(\chi_{c2} \to \gamma J/\psi)$ & - & 0.69$\pm$0.27$\pm$0.11 & - \\
& & & & \\
CLEO(1994)~\cite{cleo1} & $\Gamma_{\gamma\gamma}\times \mathcal{B}(\chi_{c2} \to \gamma J/\psi)$ & - & 0.74$\pm$0.21$\pm$0.18 & - \\
CLEO(2001)~\cite{cleo2} & $\Gamma_{\gamma\gamma}\times \mathcal{B}(\chi_{cJ} \to \gamma J/\psi)$ & 3.09$\pm$0.54$\pm$0.44 & 0.51$\pm$0.14$\pm$0.09 & 0.17$\pm$0.06$\pm$0.04 \\
CLEO(2006)~\cite{cleo3} & $\Gamma_{\gamma\gamma}\times \mathcal{B}(\chi_{c2} \to \gamma J/\psi)$ & - & 0.55$\pm$0.06$\pm$0.05 & - \\
& & & & \\
Belle(2002)~\cite{belle1} & $\Gamma_{\gamma\gamma}\times \mathcal{B}(\chi_{c2} \to \gamma J/\psi)$ & - & 0.56$\pm$0.05$\pm$0.05 & - \\
Belle(2007)~\cite{belle2} & $\Gamma_{\gamma\gamma}\times \mathcal{B}(\chi_{cJ} \to K_{S}^{0}K_{S}^{0}$) & 2.53$\pm$0.23$\pm$0.40 & 0.46$\pm$0.08$\pm$0.09 & 0.18$\pm$0.03$\pm$0.04 \\ 
Belle(2007)~\cite{belle3}$^{**}$ & $\Gamma_{\gamma\gamma}\times \mathcal{B}(\chi_{cJ} \to 4\pi$) & 1.84$\pm$0.15$\pm$0.27 & 0.40$\pm$0.04$\pm$0.07 & 0.22$\pm$0.03$\pm$0.05 \\
& $\Gamma_{\gamma\gamma}\times \mathcal{B}(\chi_{cJ} \to 2\pi2K$) & 2.07$\pm$0.20$\pm$0.40 & 0.44$\pm$0.04$\pm$0.16 & 0.21$\pm$0.03$\pm$0.09 \\
& $\Gamma_{\gamma\gamma}\times \mathcal{B}(\chi_{cJ} \to 4K$) & 2.88$\pm$0.47$\pm$0.53 & 0.62$\pm$0.12$\pm$0.12 & 0.21$\pm$0.05$\pm$0.06 \\ 
& & & & \\
This measurement & $\mathcal{B}(\psi(2S) \to \gamma \chi_{cJ})\times \mathcal{B}_{\gamma\gamma}$ & 2.53$\pm$0.37$\pm$0.26 & 0.60$\pm$0.06$\pm$0.06 & 0.24$\pm$0.04$\pm$0.03 \\ 
& & & & \\
\multicolumn{2}{c}{{\bf Averages} (weighted by total errors)} & 2.31$\pm$0.10$\pm$0.12 & 0.51$\pm$0.02$\pm$0.02 & 0.20$\pm$0.01$\pm$0.02 \\ \hline \hline
\end{tabular}
\label{review}
\end{center}
\begin{flushleft}
$^{*}$ The first error is statistical. The second error is systematic error 
combined in quadrature with the error in the branching fractions and widths 
used. The results from the literature have been reevaluated by using the 
current PDG values for branching fractions and total widths. For these results
the errors in $\mathcal{R}$ have been evaluated without taking into account
possible correlations in the systematic errors in
$\Gamma_{\gamma\gamma}(\chi_{c0})$ and $\Gamma_{\gamma\gamma}(\chi_{c2})$. \\
$^{**}$ The Belle publication gives only the product branching fractions
$\Gamma_{\gamma\gamma} \times \mathcal{B}(\chi_{c0,c2} \to \mathrm{hadrons})$.
We have calculated $\Gamma_{\gamma\gamma}$ and $\mathcal{R}$ by using the 
PDG 07~\cite{pdg07} values of branching fractions for the individual decays.
\end{flushleft}
\end{table*}

By requiring an additional resonance in the spectrum of Fig.~1 corresponding 
to $\chi_{c1}(^{3}P_{1})$, whose two-photon decay is forbidden by the 
Landau-Yang theorem~\cite{landau}, we obtain the 90$\%$ confidence limit
$\mathcal{B}(\chi_{c1} \to \gamma\gamma)<3.5\times 10^{-5}$, which is
nearly two orders of magnitude lower than the present limit quoted in
PDG 07~\cite{pdg07}. It corresponds to the 90$\%$ confidence limit
$\Gamma_{\gamma\gamma}(\chi_{c1})<0.03$ keV.

Our final results are compared to those of previous measurements in
Table~\ref{review}.
As mentioned earlier, most of the results for
$\Gamma_{\gamma\gamma}(\chi_{cJ})$ in Table~\ref{review} are from measurements
of the \textit{formation} of $\chi_{cJ}$ in two-photon fusion. 
The results listed in Table~\ref{review} for
$\Gamma_{\gamma\gamma}(\chi_{cJ})$ have been updated by using the current 
PDG~\cite{pdg07} values for the branching fractions and widths required for 
evaluating $\Gamma_{\gamma\gamma}(\chi_{cJ})$ from the directly measured
quantities.

From Table~\ref{review} we notice that although the individual results for
$\Gamma_{\gamma\gamma}(\chi_{c0})$ and $\Gamma_{\gamma\gamma}(\chi_{c2})$
show large variations, their average agrees with our results. 

We note in Table~\ref{review} that there are few 
simultaneous measurements of $\Gamma_{\gamma\gamma}(\chi_{c0})$ and
$\Gamma_{\gamma\gamma}(\chi_{c2})$, and even in those few cases $\mathcal{R}$
is seldom calculated. To put our result for $\mathcal{R}$ in perspective, we 
have calculated $\mathcal{R}$ for all simultaneous measurements in
Table~\ref{review}. A weighted average of these is $\mathcal{R}$=0.20$\pm$0.03
which is in good agreement with our result, $\mathcal{R}$=0.24$\pm$0.06.

The theoretical pQCD prediction for $\mathcal{R}$ is~\cite{barbieri}
\begin{eqnarray}
\mathcal{R}_{th}&=&(4/15)\left[1-1.76\alpha_{s}\right],~~\mathrm{since} \nonumber \\
\Gamma_{\gamma\gamma}(\chi_{c2})&=&4(|\Psi^{\prime}(0)|^{2}\alpha_{em}^{2}/m_{c}^{4})\times\left[1-1.70\alpha_{s}\right], \nonumber \\
\Gamma_{\gamma\gamma}(\chi_{c0})&=&15(|\Psi^{\prime}(0)|^{2}\alpha_{em}^{2}/m_{c}^{4})\times\left[1+0.06\alpha_{s}\right]. \nonumber 
\end{eqnarray}
where the quantities in the square brackets are the first order radiative
correction factors.

The radiative correction factor for $\Gamma_{\gamma\gamma}(\chi_{c2})$ 
(for $\alpha_{s}\simeq$0.32~\cite{pdg07}) is nearly a factor of two,
which strongly suggests possible problems with the radiative corrections.
Unfortunately, a measurement of $\Gamma_{\gamma\gamma}(\chi_{c2})$ alone
cannot provide further insight into the problem because the charm quark
mass $m_{c}$ and derivative of the wave function at origin $\Psi^{\prime}(0)$ 
are not known. However, since both unknowns cancel in the ratio 
$\mathcal{R}$, a measurement of $\mathcal{R}$ can do so, as noted, for
example, by Voloshin~\cite{voloshin}.
For $\alpha_{s}$=0.32, the predicted value, which only depends
on radiative corrections, is $\mathcal{R}_{\mathrm{th}}$=0.12. Our 
experimental result,
$\mathcal{R}_{\mathrm{exp}}$=0.24$\pm$0.06, together with the other 
determinations of $\mathcal{R}$ in Table~\ref{review} leads to the average 
$\langle \mathcal{R}_{\mathrm{exp}}\rangle$=0.20$\pm$0.02. This result 
provides experimental confirmation of the inadequacy of the present 
first-order radiative corrections, which have been often used to make 
theoretical predictions of $\Gamma_{\gamma\gamma}(\chi_{cJ})$ and 
experimental derivations of $\alpha_{s}$.

The above experimental results for $\mathcal{R}$ emphasize the need
for calculations of radiative corrections to higher orders. 
Alternatively, as noted by Buchm\"{u}ller~\cite{buchmueller}, a different 
choice of the renormalization scheme and renormalization scale should be
considered in order to arrive at a more convergent way of specifying the 
radiative corrections.

We gratefully acknowledge the effort of the CESR staff in providing us with 
excellent luminosity and running conditions. This work was supported by
the A.P.~Sloan Foundation, the National Science Foundation, the U.S. 
Department of Energy, the Natural Sciences and Engineering Research 
Council of Canada, and the U.K. Science and Technology Facilities Council.

\end{document}